\documentclass[superscriptaddress,twocolumn,amsmath,amssymb,showpacs,floatfix,prb]{revtex4}

\usepackage{graphicx}
\usepackage{dcolumn}
\usepackage{bm}
\usepackage{rotating}
\usepackage{booktabs}
\usepackage{color}

\newcommand{\tb}{Tb$_2$Ti$_2$O$_7$}
\newcommand{\gd}{Gd$_2$Ti$_2$O$_7$}
\newcommand{\R}{R$_2$Ti$_2$O$_7$}

\newcommand{\ho}{Ho$_2$Ti$_2$O$_7$}
\newcommand{\dy}{Dy$_2$Ti$_2$O$_7$}

\begin{document}

\title{Phonon and crystal field excitations in geometrically frustrated rare earth titanates}

\author{T.T.A. Lummen}
\affiliation{Zernike Institute for Advanced Materials, University of
Groningen, Nijenborgh 4, 9747 AG Groningen, The Netherlands}
\date{\today}
\author{I.P. Handayani}
\affiliation{Zernike Institute for Advanced Materials, University of
Groningen, Nijenborgh 4, 9747 AG Groningen, The Netherlands}
\author{M.C. Donker}
\affiliation{Zernike Institute for Advanced Materials, University of
Groningen, Nijenborgh 4, 9747 AG Groningen, The Netherlands}
\author{D. Fausti}
\affiliation{Zernike Institute for Advanced Materials, University of
Groningen, Nijenborgh 4, 9747 AG Groningen, The Netherlands}
\date{\today}
\author{G. Dhalenne}
\affiliation{Laboratoire de Physico-Chimie de l'Etat Solide, CNRS,
UMR8182, Universit\'{e} Paris-Sud, B\^{a}timent 414, 91405 Orsay,
France}
\author{P. Berthet}
\affiliation{Laboratoire de Physico-Chimie de l'Etat Solide, CNRS,
UMR8182, Universit\'{e} Paris-Sud, B\^{a}timent 414, 91405 Orsay,
France}
\author{A. Revcolevschi}
\affiliation{Laboratoire de Physico-Chimie de l'Etat Solide, CNRS,
UMR8182, Universit\'{e} Paris-Sud, B\^{a}timent 414, 91405 Orsay,
France}
\author{P.H.M. van Loosdrecht}
\email{P.H.M.van.Loosdrecht@rug.nl}
\affiliation{Zernike Institute
for Advanced Materials, University of Groningen, Nijenborgh 4, 9747
AG Groningen, The Netherlands}
\date{\today}

\begin{abstract}
The phonon and crystal field excitations in several rare earth
titanate pyrochlores are investigated. Magnetic measurements on
single crystals of \gd, \tb, \dy\ and \ho\ are used for
characterization, while Raman spectroscopy and terahertz time domain
spectroscopy are employed to probe the excitations of the materials.
The lattice excitations are found to be analogous across the
compounds over the whole temperature range investigated (295-4 K).
The resulting full phononic characterization of the \R\ pyrochlore
structure is then used to identify crystal field excitations
observed in the materials. Several crystal field excitations have
been observed in \tb\ in Raman spectroscopy for the first time,
among which all of the previously reported excitations. The presence
of additional crystal field excitations, however, suggests the
presence of two inequivalent Tb$^{3+}$ sites in the low temperature
structure. Furthermore, the crystal field level at approximately 13
cm$^{-1}$ is found to be both Raman and dipole active, indicating
broken inversion symmetry in the system and thus undermining its
current symmetry interpretation. In addition, evidence is found for
a significant crystal field-phonon coupling in \tb. These findings
call for a careful reassessment of the low temperature structure of
\tb, which may serve to improve its theoretical understanding.
\end{abstract}


\maketitle

\section{Introduction}
The term 'geometrical frustration'\cite{ram94,schi96,diep94} applies
to a system, when it is unable to simultaneously minimize all of its
magnetic exchange interactions, solely due to its geometry.
Magnetically interacting spins residing on such lattices are unable
to order into a unique magnetic ground state due to the competing
magnetic interactions between different lattice sites. Instead of
selecting a single, unique magnetic ground state at low
temperatures, a pure magnetically frustrated system has a
macroscopically degenerate ground state. In real systems however,
any secondary, smaller term (arising from single-ion or exchange
anisotropy, further neighbor interactions, dipolar interactions,
small lattice distortions or a magnetic field, for example) in the
system's Hamiltonian can favor certain magnetic ground states at
very low temperatures, thereby (partially) lifting this peculiar
degeneracy. In this fact lies the origin of the vast richness and
diversity of the low temperature magnetic behavior of different
frustrated systems in nature\cite{gree06,gree01,col97,ram94,moe01}.

Geometries suitable to exhibit frustration typically consist of
infinite networks of triangles or tetrahedra, which share one or
more lattice sites. One of the most common structures known to be
able to induce magnetic frustration is that of the pyrochlores,
A$_{2}$B$_{2}$O$_{7}$, where both the A$^{3+}$ ions (rare earth
element, coordinated to 8 O atoms) and the B$^{4+}$ ions (transition
metal element, coordinated to 6 O atoms) reside on a lattice of
corner-sharing tetrahedra, known as the pyrochlore lattice. Thus, if
either A$^{3+}$ or B$^{4+}$ is a magnetic species, frustration may
occur due to competing interactions. A subclass of the pyrochlores
is formed by the rare earth titanate family, \R, where the R$^{3+}$
ion is the only (para-)magnetic species, since Ti$^{4+}$ is
diamagnetic ($3d^{0}$). For the pyrochlore lattice, both
theory\cite{vil79,rei91,moe98} and Monte Carlo
simulations\cite{moe98,rei92} predict a 'collective paramagnetic'
ground state, or the lack of long range magnetic ordering, for
classical Heisenberg spins at finite temperature. The quantum
Heisenberg spin (S = 1/2) model for the pyrochlore lattice also
predicts a quantum disordered system at finite temperatures, a state
often referred to as a 'spin liquid'\cite{can98}. However, in
reality the different perturbative terms in the corresponding
Hamiltonian result in quite diverse low temperature magnetic
behavior among the different rare earth titanates\cite{gree06}, of
which the Gd, Tb, Ho and Dy variants are studied here.

The supposedly least complex case is that of gadolinium titanate,
\gd. The Gd$^{3+}$ ion has, in contrast to the Tb$^{3+}$, Ho$^{3+}$
and Dy$^{3+}$ ions, a spin only $^{8}S_{7/2}$ ($L = 0$) ground
state, rendering the influence of crystal field levels and possible
induced Ising-like anisotropy insignificant in \gd. The
experimentally determined Curie-Weiss temperature of \gd\ is $\simeq
-10$ K\cite{raju99,cash68,ram02}, indicating antiferromagnetic
nearest neighbor interactions. Thus, \gd\ could be considered as an
ideal realization of the frustrated Heisenberg antiferromagnet with
dipolar interactions. Experimentally, \gd\ has been found to undergo
a magnetic ordering transition at $\simeq$ 1 K\cite{raju99}.
However, this transition corresponds to only partial ordering of the
magnetic structure, as only 3 spins per tetrahedron
order\cite{cham01}. In this partially ordered state the spins
residing on the [111] planes of the crystal (which can be viewed as
Kagom\'{e} planes) are ordered in a $120^{\circ}$ configuration,
parallel to the Kagom\'{e} plane, while the spins residing on the
interstitial sites remain either statically or dynamically
disordered \cite{stew04}. Subsequent experimental investigations
revealed a second ordering transition at $\simeq$ 0.7 K,
corresponding to the partial ordering of the interstitial disordered
spins\cite{stew04,ram02} which, however, do remain (partially)
dynamic down to 20mK.\cite{yao05,dun06} Despite of \gd\ supposedly
being well approximated by the Heisenberg antiferromagnet with
dipolar interactions, theoretical justification for this complex
magnetic behavior remains difficult\cite{raju99,pal00,yao05}.

In \tb, the dominant interactions are antiferromagnetic, as
indicated by the experimentally determined Curie-Weiss temperature,
$\;\theta_{CW}$ $\simeq -19$ K\cite{gar99}. A study of the diluted
compound (Tb$_{0.02}$Y$_{0.98}$)$_2$Ti$_2$O$_7$ revealed that the
contribution to $\;\theta_{CW}$ due to exchange and dipolar
interactions is $\simeq -13$ K, comparable to the
$\;\theta_{CW}$-value found in \gd\cite{gin00}. Despite the energy
scale of these interactions, the Tb$^{3+}$ moments do not show long
range magnetic order down to as low as 50 mK, making it the system
closest to a real 3D \emph{spin liquid} to date\cite{gar99,gar03}.
However, crystal field (CF) calculations indicate a ground state
doublet and Ising-like easy axis anisotropy for the ($^{7}F_{6}$)
Tb$^{3+}$ magnetic moments along their local $<\!\!111\!\!>$
directions (the direction towards the center of the tetrahedron the
particular atom is in), which would dramatically reduce the degree
of frustration in the
system\cite{gin00,ros00,mirebeau07,malkin04,gard01}. Theoretical
models taking this anisotropy into account predict magnetic ordering
temperatures of about 1 K\cite{gin00,dHe00}. Subsequent theoretical
work suggests that the magnetic moment anisotropy is more isotropic
than Ising-like, which could suppress magnetic ordering\cite{kao03}.
Recently, \tb\ was argued to be in a quantum mechanically
fluctuating 'spin ice' state\cite{ice,mol07}. Virtual quantum
mechanical CF excitations (the first excited CF doublet is separated
by only $\simeq 13$ cm$^{-1}$ from the ground state
doublet\cite{gin00,mirebeau07,gard01}) are proposed to rescale the
effective theoretical model from the unfrustrated Ising
antiferromagnet to a frustrated \emph{resonating spin ice} model.
Nevertheless, the experimentally observed lack of magnetic ordering
down to the millikelvin range and the true magnetic ground state in
\tb\ remain as of yet enigmatic\cite{enj04,mol07,mir06,curnoe07}.

Illustrating the diversity in magnetic behavior due to the subtle
differences in the rare earth species of the titanates, the
situation in both \dy\cite{ram99,sny04,ram00} and
\ho\cite{har97,bra01,cor01,pet03} is again different. The R$^{3+}$
ions in these compounds have a $^{6}H_{15/2}$ (Dy$^{3+}$) and a
$^{5}I_{8}$ (Ho$^{3+}$) ground state, respectively, with
corresponding free ion magnetic moments $\mu = 10.65 \; \mu_{B}$
(Dy$^{3+}$) and $\mu = 10.61 \; \mu_{B}$(Ho$^{3+}$). These systems
were first thought to have weak ferromagnetic nearest neighbor
exchange interactions, as indicated by the small positive values of
$\theta_{CW}$, $\simeq 2 \;$ K and $\simeq 1 \;$ K for \dy\ and \ho,
respectively \cite{bram00}. More recently, however, the nearest
neighbor \emph{exchange} interactions in \dy\ and \ho\ were argued
to be \emph{anti}ferromagnetic \cite{mel04}. The effective
ferromagnetic interaction between the spins in fact is shown to be
due to the dominant ferromagnetic long-range magnetic dipole-dipole
interactions\cite{dHe00,bram01}. The R$^{3+}$-ions in both \dy\ and
\ho\ are well described by a well separated Ising doublet (first
excited states are at $\sim 266$ and $\sim 165$ cm$^{-1}$
,respectively\cite{ros00}) with a strong single ion anisotropy along
the local $<\!\!111\!\!>$ directions. Unlike for
antiferromagnetically interacting spins with local $<\!\!111\!\!>$
Ising anisotropy, \emph{ferro}magnetically interacting Ising spins
on a pyrochlore lattice should be highly
frustrated\cite{har97,bram98}. As Anderson already pointed out half
a century ago\cite{and56}, the resulting model is analogous to
Pauling's ice model\cite{paul35}, which earned both \dy\ and \ho\
the title of 'spin ice' compound\cite{har97,bram98,bram01,isa05}.
Although numerical simulations predict long range order at low
temperatures for this model\cite{mel04}, experimental studies report
no transition to a long range ordered state for either
\dy\cite{dHe00,ram99,fuk02,fen02} or \ho\cite{har97,har98,ehl04},
down to as low as 50 mK.

As is apparent from above considerations, the low temperature
magnetic behavior of the rare earth titanates is dictated by the
smallest of details in the structure and interactions of the
material. Therefore, a comprehensive experimental study of the
structural, crystal field and magnetic properties of these systems
may serve to clarify unanswered questions in their understanding. In
this paper, dc magnetic susceptibility measurements, polarized Raman
scattering experiments and terahertz time domain spectroscopy on
aforementioned members of the rare earth titanates family are
employed to gain more insight into the details that drive them
towards such diverse behavior. Raman scattering allows for
simultaneous investigation of structural and crystal field (CF)
properties through the observation of both phononic and CF
excitations, while the comparison between the various members helps
identify the nature of the different excitations observed.

\section{Experimental}
\subsection{Sample Preparation}
Polycrystalline samples of \R\ (where R = Gd, Tb, Dy, Ho) were
synthesized by firing stoichiometric amounts of high purity
($>99.9\%$) TiO$_{2}$ and the appropriate rare earth oxide
(Gd$_{2}$O$_{3}$, Tb$_{4}$O$_{7}$, Dy$_{2}$O$_{3}$ or
Ho$_{2}$O$_{3}$, respectively), in air, for several days with
intermittent grindings. The resulting polycrystalline powder was
subsequently prepared for single crystal growth, using the method
described by Gardner, Gaulin and Paul\cite{gar98}. The following
single crystal growth (also done as described in ref. 49) using the
floating zone technique yielded large, high quality single crystals
of all of the \R\ variants. Discs ($\simeq$ 1 mm thickness) with
a,b-plane surfaces were cut from oriented single crystals and
subsequently polished, in order to optimize scattering experiments.
The \tb\ sample used in Raman experiments was subsequently polished
down to $\simeq$ 250 $\mu$m thickness to facilitate THz transmission
measurements.

\subsection{Instrumentation}
X-ray Laue diffraction, using a Philips PW 1710 diffractometer
equipped with a Polariod XR-7 system, was employed to orient the
single crystal samples of \gd\ and \tb\ for the polarized Raman
spectroscopy experiments, while simultaneously confirming the single
crystallinity of the samples. The \dy\ and \ho\ single crystals were
oriented using an Enraf Nonius CAD4 diffractometer.

The magnetic susceptibilities of the obtained rare earth titanates
were measured using the Quantum Design MPMS-5 SQUID magnetometer of
the 'Laboratoire de Physico-Chimie de l'Etat Solide' (LPCES), CNRS,
UMR8182 at the 'Universit\'{e} Paris-Sud  in Orsay, France. The \R\
samples, about 100 mg of single crystal (in the form of discs of
approximately 4 mm diameter and 1 mm thickness), were placed in
cylindrical plastic tubes and locked in position. Next, the samples
were zero-field-cooled down to 1.7 K, after which the magnetization
of the sample was measured as a function of the temperature in an
applied magnetic field of 100 Oe, while warming the sample.

Polarization controlled, inelastic light scattering experiments were
performed on all oriented \R\ samples. The experiments were
performed in a $180^{\circ}$ backscattering configuration, using a
triple grating micro-Raman spectrometer (T64000 Jobin-Yvon),
consisting of a double grating monochromator (acting as a spectral
filter) and a polychromator which disperses the scattered light onto
a liquid nitrogen cooled CCD detector. The frequency resolution was
better than 2 cm$^{-1}$ for the frequency region considered. The
samples were placed in a liquid helium cooled optical flow-cryostat
(Oxford Instruments). The temperature was stabilized with an
accuracy of 0.1 K in the whole range of measured temperatures (from
2.5 to 295 K). The 532.6 nm (frequency doubled) output of a
Nd:YVO$_{4}$ laser was focused on the \gd, \tb\ and \dy\ samples
using a 50x microscope objective and used as excitation source in
the scattering experiments. A Krypton laser (676.4 nm) was used as
the excitation source for the scattering experiments on the \ho\
sample, since 532.6 nm excitation (resonant at low temperatures in
case of \ho) results in fluorescence dominating the inelastic
scattering spectrum in the 5-800 cm$^{-1}$ spectral range. The power
density on the samples was of the order of 50 $\mu$W/$\mu$m$^{2}$ in
all cases. The polarization was controlled both on the incoming and
outgoing beam. Parallel (\emph{$\|$}) and perpendicular ($\bot$)
measurements on \gd\ and \tb\ were performed along crystallographic
axes of the a,b surface of the samples, Porto notations c(aa)c and
c(ab)c, respectively. Unfortunately, the orientation of the a and b
axes in the (a,b) plane of the \dy\ and \ho\ surfaces with respect
to the light polarizations was not known. Analogous Porto notations
are c(xx)c (\emph{$\|$}) and c(xy)c ($\bot$), respectively, where x
is a direction in the (a,b) plane of the sample making an
undetermined angle $\alpha$ with the a axis, while y, in the same
a,b plane of the crystal, is perpendicular to the x direction. Raman
spectra were fitted with lorentzian lineshapes to extract mode
parameters.

    Terahertz time domain spectroscopy (THz TDS)\cite{schmut04} was performed on \tb\
using a home made setup similar to those described
elsewhere\cite{beard02, schmut04}. THz pulses (pulse duration of
several ps, frequency range 0.3-2.5 THz) were generated through a
difference frequency generation process in a ZnTe single crystal
upon pulsed excitation (120 fs, 800 nm) by an amplified Ti:sapphire
system. The magnitude of the time dependent electric field
transmitted through the sample (w.r.t. that transmitted through
vacuum) was measured at various temperatures, through electro-optic
sampling in a second ZnTe single crystal, using 800 nm pulses of
approximately 120 fs. The sample used was a thin slice of single
crystalline \tb\ (the same sample as used in the Raman experiments),
which was mounted on a copper plate with an aperture ($\varnothing$
2 mm) and placed in a liquid helium cooled optical flow-cryostat
(Oxford Instruments). The polarization of the THz radiation was
parallel to the crystallographic a-axis.

\section{Results and Discussion}
\subsection{Magnetic Measurements} 
\begin{figure*}[htb]
\centering
\includegraphics{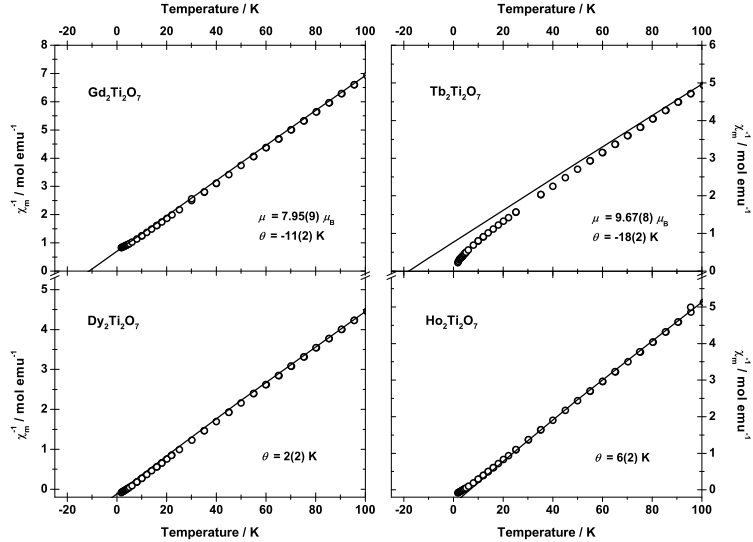}
\caption{\label{All_inversechi}Inverse molar magnetic
susceptibilities (open circles) of all \R\ samples and their
corresponding Curie-Weiss fits (solid lines, eq. \ref{CWlaw}) versus
temperature. The Curie-Weiss law, which is fitted to the higher
temperature data ($>$ 100 K) is obeyed down to low temperatures.}
\end{figure*}

The magnetic susceptibility
$\chi$, defined as the ratio of the magnetization of the sample to
the applied magnetic field, of all the rare earth titanate samples
was measured in a 100 Oe applied magnetic field. Since the samples
used were plate-like discs, the data have been corrected by a
demagnetization factor as calculated for flat, cylindrical
plates\cite{sato89}. Fig. \ref{All_inversechi} shows the inverse
molar susceptibilities of all samples in the low temperature regime.
For each sample, the data was fitted to a Curie-Weiss form for the
molar magnetic susceptibility of an antiferromagnet:

\begin{equation}
\chi_{m}=\frac{2C}{(T-\theta)}+B, \label{CWlaw}
\end{equation}
\newline
where C is the Curie constant in CGS units ($C =
N_{A}\mu^{2}\mu_{B}^{2}/3k_{b}\simeq \mu^{2}/8$, in
[emu$\cdot$K$\cdot$mol$^{-1}$]), $\theta$ is the expected transition
temperature (giving an indication of the sign of the magnetic
interactions) and $B$ is a temperature independent Van Vleck
contribution to the susceptibility. The model was fitted to high
temperature experimental data (100 K and up, where the
demagnetization correction is of the order of 1 \%) and linear
regression analysis of $\chi^{-1}_{m,i}$ yielded the experimental
values for $\theta_{i}$, $\mu_{i}$ and $B_{i}$. These are tabulated
below in table 1, together with several values reported in
literature.

\begin{table}[htb]
\caption{\label{CWparameters}Experimentally determined and
literature values for $\theta_{i}$, $\mu_{i}$ and $B_{i}$ for the
different \R\ compounds. The numbers in between brackets give the
accuracy in the last decimal of the values.}
\begin{ruledtabular}
\begin{tabular}{lcD{,}{.}{2}D{,}{.}{-1}}
\multicolumn{1}{c}{Compound \emph{i} $(source)$}    &   \multicolumn{1}{c}{$\theta_{i} (K)$}    &   \multicolumn{1}{c}{$\mu_{i} (\mu_{B})$} &   \multicolumn{1}{c}{$B_{i} (emu \cdot mol^{-1})$}\\
\hline
\gd\ (\emph{exp.})                  & -11(2)                        & 7,95(9)           & \multicolumn{1}{c}{0\footnotemark[1]}\\
\gd\ (\emph{lit.})\cite{raju99}     & -9,6                          & 7,7               & \multicolumn{1}{c}{-}\\
\gd\ (\emph{lit.})\cite{cash68}     & -11,7                         & 7,8               & \multicolumn{1}{c}{-}\\
\gd\ (\emph{lit.})\cite{bram00}     & -8,95(6)                      & 7,224(3)          & 6,0 \cdot 10^{-4}\\
\tb\ (\emph{exp.})                  & -18(2)                        & 9,67(7)           & 3,4(9) \cdot 10^{-3} \\
\tb\ (\emph{lit.})\cite{gar99,gin00}&  -19                          & 9,6               & \multicolumn{1}{c}{-}\\
\dy\ (\emph{exp.})                  & 2(2)                        & \multicolumn{1}{c}{-\footnotemark[2]} & \multicolumn{1}{c}{-\footnotemark[2]}\\
\dy\ (\emph{lit.})\cite{bram00}     & $\simeq$ 1,0                 & 9,590(6)          & 1,1(1)\cdot 10^{-2}\\
\ho\ (\emph{exp.})                  & 6(2)                        & \multicolumn{1}{c}{-\footnotemark[2]} & \multicolumn{1}{c}{-\footnotemark[2]}\\
\ho\ (\emph{lit.})\cite{bram00}     & $\simeq$ 2,0                  & 9,15(3)           & 1,2(1) \cdot 10^{-2}\\
\bottomrule
\end{tabular}
\end{ruledtabular}
\footnotetext[1]{When fitting $\chi_{m}$ with a nonzero \emph{B}
term, it becomes negligibly small. The best fit is obtained with
\emph{B=0}.} \footnotetext[2]{Since dipolar interactions are
dominant in this case, no reliable values were extracted (see
text).}
\end{table}

In general, the experimentally obtained data compare (where
possible) favorable to the various values reported in literature
(table \ref{CWparameters}). The extracted $\theta$ parameters for
\dy\ and \ho\ do slightly deviate from literature values, presumably
due to the estimation of the demagnetization factor (fits to the
present uncorrected data yield $\theta$ values of approximately 1
and 2 K, respectively). The experimentally determined paramagnetic
moments obtained for \gd\ and \tb\ are also in excellent agreement
with the corresponding free ion values, which are $\mu$ = 7.94
$\mu_{B}$ and $\mu$ = 9.72 $\mu_{B}$ for the Gd$^{3+}$
($^{8}S_{7/2}$) and Tb$^{3+}$ ($^{7}F_{6}$) free ions, respectively.
The large negative Curie-Weiss temperatures for \gd\ and \tb\
indicate antiferromagnetic exchange coupling. In contrast, the
small, positive $\theta$ values for \ho\ and \dy\ initially led to
the assumption of weak ferromagnetic exchange interactions between
nearest neighbor Dy$^{3+}$ and Ho$^{3+}$ ions\cite{bram00}. As
stated above, however, since the Ho$^{3+}$ and Dy$^{3+}$ ions have a
large magnetic moment (free ion values are $\mu$ = 10.607 $\mu_{B}$
($^{5}I_{8}$ ground state) and $\mu$ = 10.646 $\mu_{B}$
($^{6}H_{15/2}$ ground state), respectively), the dipolar
interactions between neighboring R$^{3+}$ ions are dominating the
effective nearest neighbor (n.n.) interactions. The n.n.
\emph{exchange} interactions are in fact
\emph{anti}ferromagnetic\cite{mel04}, while the dominant dipolar
n.n. interactions are of ferromagnetic nature\cite{dHe00,bram01}.
Consequently, the effective n.n. interactions are slightly
ferromagnetic, resulting in the positive $\theta$ values. Another
consequence of the dipolar interactions and being dominant is the
fact that extracting the real values of $\mu$ and $B$ from the
inverse susceptibility curves becomes non-trivial, since more
elaborate models taking the dipolar interaction into account are
needed.

\subsection{Raman Spectroscopy}
\subsubsection{Room temperature spectra}

To determine the Raman active vibrations of the single crystals,
group theory analysis was employed. This predicts that, for the
cubic rare earth titanate structure (\R) of space group
\emph{Fd\={3}m} (\emph{O$_{h}$$^{7}$}), the sublattices of the unit
cell span the following irreducible representations:
\[
\begin{array}{rrll}

16(c)$-site$: & $Ti$^{4+}-sublattice & = & A_{2u}+E_{u}+2F_{1u}+F_{2u}\\
16(d)$-site$: & $R$^{3+}-sublattice & = & A_{2u}+E_{u}+2F_{1u}+F_{2u}\\
48(f)$-site$: & $O$(1)-sublattice & = & A_{1g}+E_{g}+2F_{1g}+3F_{2g}+\\
& & & A_{2u}+E_{u}+3F_{1u}+2F_{2u}\\
8(a)$-site$: & $O$(2)-sublattice & = & F_{1u}+F_{2g}\\

\end{array}
\]

This makes the following decomposition into zone center normal modes
(excluding the F$_{1u}$ acoustic mode):
\[
\Gamma = A_{1g} + E_{g} + 2F_{1g} + 4F_{2g} + 3A_{2u} + 3E_{u} +
7F_{1u} + 4F_{2u}
\]

\begin{figure*}[htb]
\centering
\includegraphics{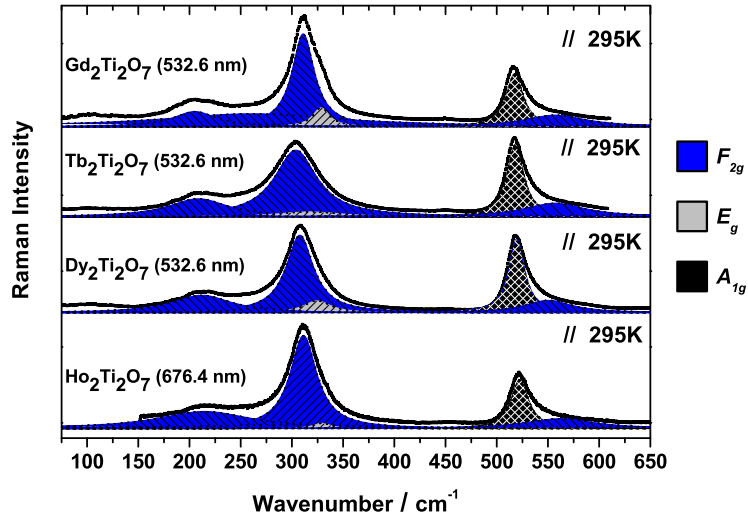}
\caption{(Color online) \label{Generalspectrum}Room temperature
Raman spectra of the \R\ crystals in parallel polarization
configuration. The assignment of the Raman active modes to the
observed peaks is indicated by the colored lorentzian lineshapes.}
\end{figure*}

Of these normal modes, only the $A_{1g}$, $E_{g}$ and the 4 $F_{2g}$
modes are Raman active. The 7 $F_{1u}$ modes are infrared active and
the remaining modes are optically inactive. The symmetry coordinates
for the optically active normal modes are given by Gupta \textit{et
al.}\cite{gup01}. Based on the symmetries of the Raman active modes,
the A$_{1g}$ and E$_{g}$ modes are expected to be observed in
parallel polarization (\emph{$\|$}) spectra, while the F$_{2g}$
modes are expected in the perpendicular polarization ($\bot$)
spectra. Surprisingly, the room temperature spectra of \R\ show all
the Raman active modes in both polarizations. Angle dependent Raman
measurements (varying the angle between the incoming light
polarization and the b-axis from $-45^{\circ}$ to $45^{\circ}$)
reveal no orientation in which the theoretical selection rules are
fully obeyed. Polycrystallinity and/or crystalline disorder is
however, not believed to be the cause, since other techniques (X-ray
Laue diffraction, magnetic measurements) reveal no such signs. Also,
to our best knowledge, no Raman spectrum of any of these rare earth
titanates fulfilling their phononic selection rules has been
published. Although no full compliance of phononic selection rules
was observed, the nature of the Raman active modes was clearly
identified, through their angular dependence. Their assignment is
indicated in fig. \ref{Generalspectrum} as well as in table
\ref{Assignments}. The figure depicts the room temperature (RT),
parallel polarization spectra of the rare earth titanates and the
corresponding modes. The assignment of the mode symmetries is based
on comparison with assignments in previous works on
these\cite{saha06,zha05,mori03,hes02,gup01,van83} and
related\cite{brown03,gle01,gupta01,gupta02,van83,vandenb83}
compounds (see table \ref{Assignments}), on aforementioned
measurements (not shown) of the angular dependence of mode
intensities and on the temperature dependence of the modes (vide
infra).

\begin{sidewaystable}[htb]
\caption{\label{Assignments}Experimentally determined and literature
values of the frequencies (in cm$^{-1}$) of observed Raman modes in
\R\ at room temperature and their respective symmetry assignments
(between brackets).}
\begin{center}
    \begin{ruledtabular}
    \begin{tabular*}{0pt}{c c c c *{4}{| c} | cc | c}
    \multicolumn{4}{ c |}{\centering \underline{This work}} & \multicolumn{1}{c |}{\underline{Saha \textit{et al.}\cite{saha06}}} & \underline{Zhang \textit{et al.}\cite{zha05}} & \underline{Mori \textit{et al.}\cite{mori03}} & \underline{Hess \textit{et al.}\cite{hes02}} & \multicolumn{2}{c |}{\centering \underline{Gupta \textit{et al.}\cite{gup01}}} & \underline{Vandenborre \textit{et al.}\cite{van83}}\\
    R = Gd  & \multicolumn{1}{c}{Tb} & \multicolumn{1}{c}{Dy} & \multicolumn{1}{c |}{Ho} & Gd & Gd  & Gd & Gd & Gd (\textit{calcd.})& Gd (\textit{obs.})& Gd \\
    \hline
    $\sim$104\footnotemark[1]   &    $\sim$102\footnotemark[1]  &  $\sim$103\footnotemark[1]  &   $\sim$105\footnotemark[1] (-)                 &    - (-)                   &   - (-)                                     & 105 (\textit{F$_{2g}$})     &   - (-)                   &   - (-)                      &   -  (-)                 & 110 (-)                \\
    $\sim$128\footnotemark[1]   &    -  &  $\sim$124\footnotemark[1]  &   $\sim$122\footnotemark[1] (-)                 &    - (-)                   &   - (-)                                     & 125 (\footnotemark[2])      &   - (-)                   &   - (-)                      &   -  (-)                 & 215 (-)                \\
    205 &  209  & 212 & 214 (\textit{F$_{2g}$}) &    215 (\textit{F$_{2g}$}) &   210 (\textit{F$_{2g}$})                   & 222 (\textit{F$_{2g}$})     &   219 (\textit{F$_{2g}$}) &   230.6 (\textit{F$_{2g}$})  &   225 (\textit{F$_{2g}$}) & 225 (\textit{F$_{2g}$})\\
    260 & 256 & 269 & 297 (\textit{F$_{2g}$}) & - (-) & - (-) & - (-) & - (-) & - (-) & - (-) & - (-) \\
    310 &  303  & 308 & 311 (\textit{F$_{2g}$})  &    312 (\textit{F$_{2g}$}) &   310\footnotemark[4] (\textit{E$_{g}$})    & 311 (\textit{E$_{g}$})      &   312 (\textit{E$_{g}$})  &   318.0 (\textit{E$_{g}$})   &   - (-)                  & 317 (\textit{F$_{2g}$})\\
    325 &  313  & 328 & 329 (\textit{E$_{g}$}) &    330 (\textit{E$_{g}$})  &   310\footnotemark[4] (\textit{F$_{2g}$})   &   -         (-)             &   - (-)                   &   328.2 (\textit{F$_{2g}$})  &   317 (\textit{F$_{2g}$}) & 347\footnotemark[5] (\textit{E$_{g}$}) \\
    450\footnotemark[3] &  450\footnotemark[3]  & 451\footnotemark[3] & 452\footnotemark[3]  (-)&    - (-)                   &   470 (\textit{F$_{2g}$})                   & 452         (-)             &   455 (\textit{F$_{2g}$}) &   - (-)                      &   - (-)                  &   - (-)                               \\
    -   &    -  &  -  &   - (-)                 &    - (-)                   &   - (-)                                     &   -         (-)             &   - (-)                   &   522.2 (\textit{F$_{2g}$})  &   -  (-)                 &   - (-)                \\
    517 &  518  & 519 & 522 (\textit{A$_{1g}$}) &    518 (\textit{A$_{1g}$}/\textit{F$_{2g}$}) &   520 (\textit{A$_{1g}$})                   & 518 (\textit{F$_{2g}$})     &   519 (\textit{A$_{1g}$}) &   526.8 (\textit{A$_{1g}$})  &   515 (\textit{A$_{1g}$})& 515 (\textit{A$_{1g}$}+\textit{F$_{2g}$}) \\
    554 &  557  & 550 & 562 (\textit{F$_{2g}$}) &    547 (\textit{F$_{2g}$}) &   570 (\textit{F$_{2g}$})                   & 585 (\textit{F$_{2g}$})     &   549 (\textit{F$_{2g}$}) &   594.0 (\textit{F$_{2g}$})  &   580 (\textit{F$_{2g}$})& 580 (\textit{F$_{2g}$})\\
    $\sim$677\footnotemark[6]  &    $\sim$689\footnotemark[6]  &  $\sim$693\footnotemark[6]  &   $\sim$701\footnotemark[6] (-) &    684 (-)                   &   - (-)                                     &   -         (-)             &   680 (\textit{F$_{2g}$}) &   - (-)  &   -  (-)                 &   - (-) \\
    $\sim$703\footnotemark[6]  &    $\sim$706\footnotemark[6]  &  $\sim$720\footnotemark[6]  &   $\sim$724\footnotemark[6] (-) &    701 (-)                   &   - (-)                                     &   -         (-)             &   - (-)                   &   - (-)  &   -  (-)                 &   705 (-) \\
    \end{tabular*}
    \end{ruledtabular}
    \footnotetext[1]{These modes are very weak and barely resolved, therefore their exact position is estimated.}
    \footnotetext[2]{Mori \textit{et al.} ascribe this band to trace amounts of Gd$_{2}$O$_{3}$.}
    \footnotetext[3]{This is a very weak mode, barely resolved above the noise. Mori \textit{et al.} ascribe this mode to trace amounts of TiO$_{2}$.}
    \footnotetext[4]{Zhang \textit{et al.} indicate the lower and higher wavenumber components of this band as the E$_{g}$ and F$_{2g}$ modes, respectively.}
    \footnotetext[5]{This band has been calculated, rather than observed by Vandenborre \textit{et al.}}
    \footnotetext[6]{These overlapping modes comprise a weak band, in which the two modes cannot be separately resolved.}
\end{center}
\end{sidewaystable}

Comparison between the RT Raman spectra yields the conclusion that
the nature of the R$^{3+}$ ion has only slight influence on the
Raman active vibrational modes. This is not surprising, since in all
the Raman active modes only the oxygen atoms are
displaced\cite{saha06,gup01}. Consequently, there is no obvious
systematic variation of the phonon frequencies with the mass of the
respective rare earth ions, as has been noted before for the rare
earth titanates\cite{gup01}, hafnates\cite{gupta02},
manganates\cite{brown03} and stannates\cite{gupta01}. The assignment
of the \R\ modes in literature has been mostly consistent (see Table
\ref{Assignments}), yet there are a few debated details. There is
general agreement on the nature of the modes at $\simeq$ 210
cm$^{-1}$ (\textit{F$_{2g}$}, O(2)-sublattice mode\cite{saha06}),
$\simeq$ 519 cm$^{-1}$ (\textit{A$_{1g}$}, R-O stretching
mode\cite{mori03,zha05}) and $\simeq$ 556 cm$^{-1}$
(\textit{F$_{2g}$}, O(1)-sublattice mode\cite{saha06}). Temperature
and angle dependent Raman measurements show the band around $\simeq$
315 cm$^{-1}$ to consist of two modes, a \textit{F$_{2g}$} mode
around 310 cm$^{-1}$ (O-R-O bending mode\cite{mori03,zha05}) and an
\textit{E$_{g}$} mode around 327 cm$^{-1}$ (O(1)-sublattice
mode\cite{saha06}), as recognized by Saha \textit{et
al.}\cite{saha06} and Vandenborre \textit{et al.}\cite{vandenb83}.
Earlier works either interchanged the mode assignment within this
band\cite{zha05,gup01} or ascribed the whole band to only one of
these modes\cite{mori03,hes02}. However, our temperature and angle
dependent measurements confirm the assignment made by Saha
\textit{et al.} The last expected phonon, an \textit{F$_{2g}$} mode,
has been either not accounted for \cite{gup01}, combined with the
\textit{A$_{1g}$} mode in one band\cite{saha06,van83} or ascribed to
low intensity peaks around 105\cite{mori03}, 450 \cite{zha05} or
680\cite{hes02} cm$^{-1}$ in previous works. Here, it is ascribed to
a broad, low intensity mode around $\simeq$ 260 cm$^{-1}$. This mode
is not clearly resolved in the RT spectra due to the fact that it
overlaps largely with the neighboring, strong \textit{F$_{2g}$}
modes at $\simeq$ 210 cm$^{-1}$ and $\simeq$ 309 cm$^{-1}$. Fitting
with those two peaks only, however, does not adequately reproduce
the experimental spectral shape in the the 200-300 cm$^{-1}$ window.
Additionally, as the temperature is lowered, the phonon modes
sharpen and the existence of this excitation becomes obvious in the
spectra. Worth noting are also the two anomalous modes in \tb\ at
$\simeq$ 303 (\textit{F$_{2g}$}) and $\simeq$ 313 cm$^{-1}$
(\textit{E$_{g}$}), which have lower frequencies and wider
lineshapes compared to their counterparts in the other,
isostructural rare earth titanates.

Next to the expected Raman active vibrations, the spectra in this
work show some very weak scattering intensity at low wavenumbers
(first two rows in table \ref{Assignments}), which has been reported
before\cite{van83,mori03}. Vandenborre \textit{et al.}\cite{van83}
were unable to account for this intensity in their calculations,
while Mori \textit{et al.}\cite{mori03} offer the plausible
assignment to trace R$_{2}$O$_{3}$ in the system. The latter
assignment is also tentatively adopted here. Mori \textit{et al.}
also suggested the 'missing' $F_{2g}$ mode may be responsible for
some of this low wavenumber intensity. Additionally, a weak mode is
observed at 450 cm$^{-1}$ in most \R\ compounds, as was also seen
before. Zhang \textit{et al.}\cite{zha05} and Hess \textit{et
al.}\cite{hes02} ascribed the 'missing' $F_{2g}$ mode to this
feature. Alternatively, Mori \textit{et al.}\cite{mori03}
interpreted it as being due to trace amounts of starting compound
TiO$_{2}$, which is known\cite{porto67} to have a phonon at 447
cm$^{-1}$. The true origin of this mode is at present unclear.
Finally, there is some low intensity scattering intensity at higher
wavenumbers, around 700 cm$^{-1}$ (last two rows of table
\ref{Assignments}). This intensity has been observed
before\cite{mori03,van83,saha06,hes02,zha05} and is ascribed to
forbidden IR modes made active by slight, local non-stoichiometry in
the system.\cite{mori03,saha06}

\subsubsection{Temperature dependence}

\begin{figure*}[htb]
\centering
\includegraphics[width=10.0cm]{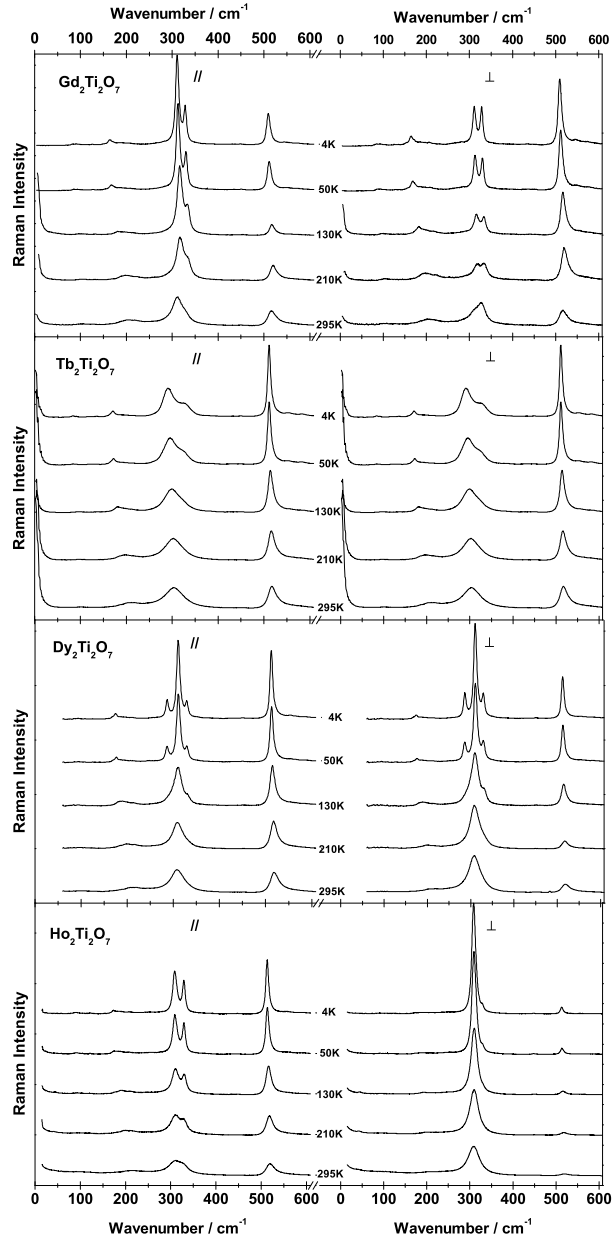}
\caption{\label{Tdependence}Temperature dependence of the \R\
parallel (\emph{$\|$}) and perpendicular ($\bot$) Raman spectra. The
spectra show several changes when going down in temperature (see
text), most of which (the phononic excitations) are analogous for
all \R\ lattices. In addition, at lower temperatures various crystal
field excitations appear in the spectra of some of the rare earth
titanates (See fig. \ref{Lowtemp}. Spectra are normalized to the
integrated intensity of the depicted frequency window.}
\end{figure*}

Raman spectra of the \R\ crystals were recorded at temperatures
ranging from RT to 4 K. Figure \ref{Tdependence} depicts the
evolution of both the parallel (\emph{$\|$}) and perpendicular
($\bot$) polarization Raman spectra of the \R\ crystals with
decreasing temperature.

\begin{figure*}[htb]
\centering
\includegraphics{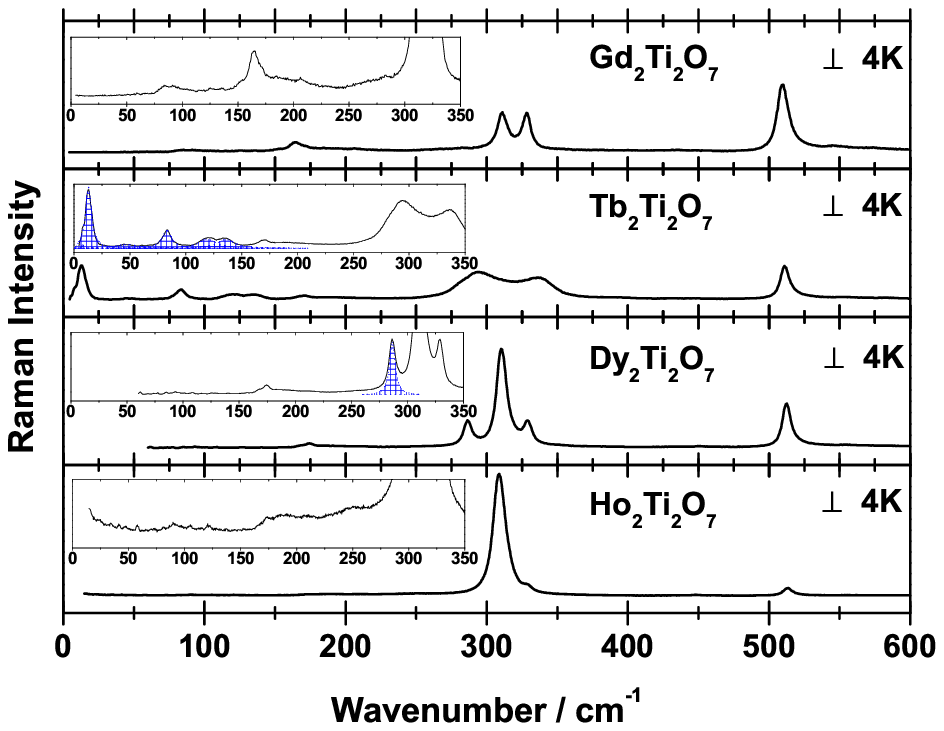}
\caption{\label{Lowtemp} (Color online) Low temperature spectra for
all \R\ as measured in perpendicular polarization configuration. The
spectra are normalized on the total integrated intensity of the
spectra. The insets show respective zoom-ins on the lower wavenumber
regions of the spectra, where the excitations identified as CF
levels of the rare earth ions are indicated by dotted blue
lorentzian peaks. \gd\ shows only phononic modes and can be regarded
as the lattice template. \tb\ and \dy\ show additional CF modes,
while \ho\ shows again only lattice excitations. A fitted lorentzian
centered at zero (Rayleigh line) is subtracted from the \tb\ data
for clarity.}
\end{figure*}

Going down in temperature several spectral changes occur in the
Raman spectra. The evolution of the phononic excitations with
temperature are very similar in the different \R\ lattices. Again,
this is not surprising, since only oxygen atoms are displaced in the
Raman active phonons\cite{saha06,gup01}. Firstly, the lowest
frequency phonon ($F_{2g}$, $\omega$ $\simeq$ 210 cm$^{-1}$ at RT)
shows a strong softening ($\omega \simeq$ 170 cm$^{-1}$ at 4 K) and
sharpening with decreasing temperature, revealing the previously
unresolved $F_{2g}$ mode ($\omega_{RT}$ $\simeq$ 260 cm$^{-1}$),
which also softens ($\omega_{4K}$ $\simeq$ 190 cm$^{-1}$), but
doesn't show a strong narrowing. Secondly, the sharpening of both
the $\simeq$ 309 cm$^{-1}$ \textit{F$_{2g}$} and the $\simeq$ 324
cm$^{-1}$ \textit{E$_{g}$} mode decreases their spectral overlap,
clearly justifying the two-mode interpretation of the $\simeq$ 315
cm$^{-1}$ band at RT. Additionally, both modes show a slight
softening upon cooldown. Also the A$_{1g}$ phonon ($\omega_{RT}$
$\simeq$ 519 cm$^{-1}$) shows the familiar softening ($\omega_{4K}$
$\simeq$ 511 cm$^{-1}$) and sharpening trend on cooling. Finally,
due to its large width and low intensity, describing the temperature
evolution of the highest frequency \textit{F$_{2g}$} phonon proves
rather difficult, though it seems to soften slightly. Comparison of
the \R\ spectra in fig. \ref{Tdependence} yields the observation
that the anomalous phonons (at $\simeq$ 303 (\textit{F$_{2g}$}) and
$\simeq$ 313 cm$^{-1}$ (\textit{E$_{g}$})) in \tb\ remain wide
throughout the temperature range, in contrast to the corresponding
modes in the other titanates. Additionally, these modes are shifting
in opposite directions in \tb\ only: the \textit{F$_{2g}$} mode
softens ($\omega_{4K}$ $\simeq$ 295 cm$^{-1}$) while the
\textit{E$_{g}$} mode considerably hardens ($\omega_{4K}$ $\simeq$
335 cm$^{-1}$). An explanation for this anomalous behavior could be
coupling of these phonons to low frequency crystal field excitations
of the Tb$^{3+}$-ions (vide infra).

\subsubsection{Crystal field excitations}

Aside from the phonons in the Raman spectra of \R, several spectra
also show crystal field (CF) excitations of the R$^{3+}$ ions at low
temperatures, as shown in fig. \ref{Lowtemp}. The CF level splitting
in the different rare earth ions depends on their electronic
configuration and their local surroundings. In the \R\ family, the
simplest case is that of the Gd$^{3+}$ ion (4$f^{7}$), which has a
spin only $^{8}S_{7/2}$ ($L = 0$) ground state, resulting in the
absence of a level splitting due to the local crystal field.
Consequently, the Raman spectrum of \gd\ shows no CF excitations,
making it a suitable 'template' of the \R\ Raman spectrum with
lattice excitations only. Combined with the strong correspondence of
the phononic excitations in the \R\ spectra, it allows for quick
identification of CF modes in the other compounds.

\begin{figure*}[htb]
\centering
\includegraphics[width=15.0cm]{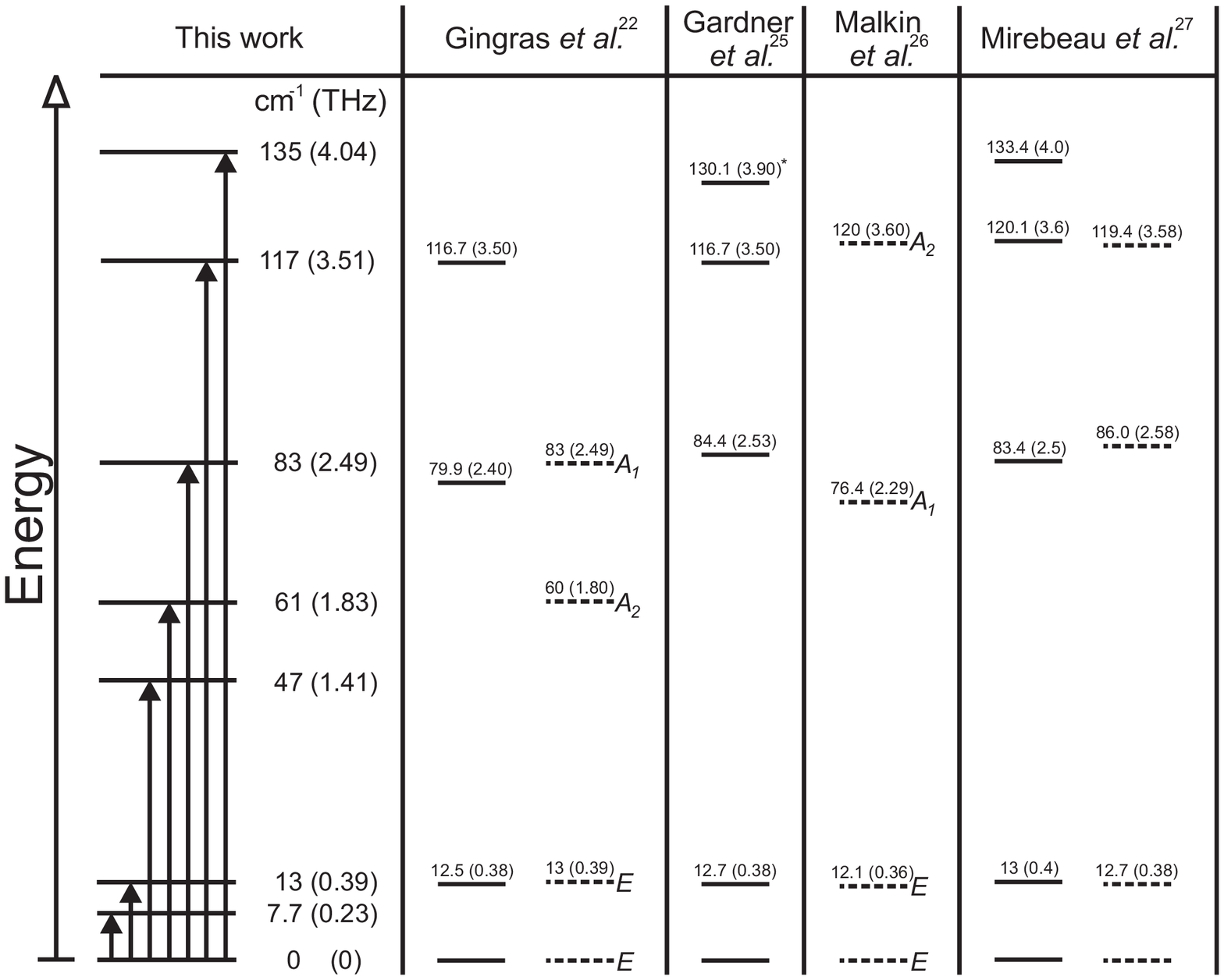}
\caption{\label{levelscheme}Various CF energy level schemes of the
Tb$^{3+}$ ion in \tb. Solid lines depict CF levels that have been
observed experimentally, while dotted lines represent CF levels as
calculated using CF calculations. Numbers indicate CF level energies
in cm$^{-1}$ (THz) and symbols in italics indicate symmetries of
corresponding levels. Solid arrows indicate all CF excitations
observed in the low temperature Raman spectrum of \tb\ (see fig.
\ref{Tb_zoom}).}
\end{figure*}

\begin{figure*}[htb]
\centering
\includegraphics{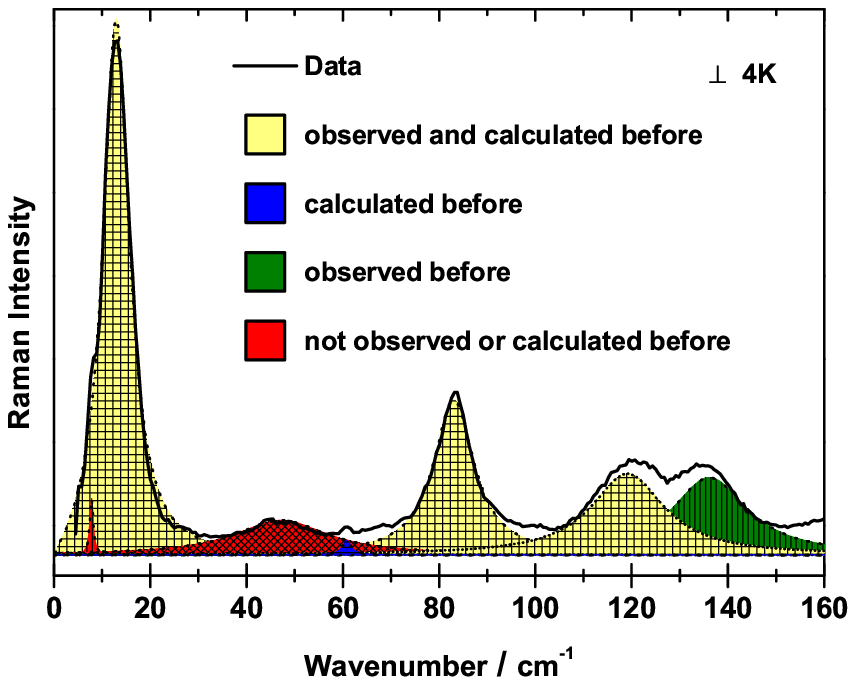}
\caption{\label{Tb_zoom}(Color online) The low wavenumber region of
the perpendicular polarization Raman spectrum of \tb\ at 4 K. A
fitted lorentzian centered at zero (Rayleigh line) is subtracted
from the data for clarity. The crystal field levels (see fig.
\ref{levelscheme}) are indicated by the filled lorentzian lineshapes
that were fit to the data. A distinction has been made between CF
levels that have been calculated and observed before (yellow), CF
levels that have been calculated before, but so far have not been
observed (blue), CF levels that have been observed before, but have
not been calculated (green) and additional CF levels that are
observed for the first time using inelastic light scattering (red).}
\end{figure*}

More complicated is the CF level splitting in the Tb$^{3+}$ ion
(4$f^{8}$), which has a $^{7}F_{6}$ ground state. Several studies
calculating the crystal field for the Tb$^{3+}$-ion in
\tb\cite{gin00,mirebeau07,gard01,malkin04}, based on inelastic
neutron scattering
results\cite{gin00,gard01,gau98,kan99,gar99,mirebeau07}, yielded
slightly differing energy level schemes for the lowest crystal field
levels of Tb$^{3+}$ in \tb\, as is schematically depicted in fig.
\ref{levelscheme}. Here, solid lines depict CF levels observed
experimentally, while dotted lines indicate CF levels obtained
through CF calculations. Shown in fig. \ref{Tb_zoom}, which is a
zoom-in on the low-wavenumber region of the 4 K perpendicular
polarization spectrum of \tb, are the CF excitations that are
observed using inelastic light scattering. As is also clear from
fig. \ref{levelscheme}, all CF excitations previously observed using
inelastic neutron scattering are also observed here. Furthermore,
additional low-lying excitations can be seen, which are easily
identified as CF levels, by comparison with the 'lattice-only
template' spectrum of \gd.

The excitations from the crystal field ground state to the higher
crystal field levels, of approximate calculated values 13, 60, 83
and 118 cm$^{-1}$ (see fig. \ref{levelscheme}), are all observed at
very similar frequencies, at 12.9, 60.8, 83.1 and 119.2 cm$^{-1}$,
respectively. The 60.8 cm$^{-1}$ level has not been observed
experimentally before, though it did follow from the CF calculation
made by Gingras \emph{et al.}\cite{gin00}. Conversely, the $\simeq$
135.2 cm$^{-1}$ mode has been observed through inelastic neutron
scattering, yet has not been accounted for in CF calculations.
Although Gardner \emph{et al.}\cite{gard01} interpreted it as an
optical phonon, this excitation is clearly identified as a CF
excitation here, through the isostructural comparison with the other
titanates. The latter assignment is also made by Mirebeau \emph{et
al.}\cite{mirebeau07}. Additionally, new CF excitations are observed
at 7.7 and 47.2 cm$^{-1}$. While the former is recognized as a CF
excitation from the CF ground state to a new CF energy level, the
47.2 cm$^{-1}$ could also be interpreted as an excitation from the
excited CF level at 12.9 cm$^{-1}$ to the higher excited CF level at
60.8 cm$^{-1}$. While such an excitation is Raman active (see
below), its occurrence at 4 K seems unlikely, since at this
temperature the excited CF level (12.9 cm$^{-1}$ $\simeq$ 19 K) is
not expected to be populated enough to give rise to a measurable
Raman signal. Additionally, were it an excitation from an excited
state, its intensity would decrease upon cooling. Instead, the
intensity of this mode steadily increases with decreasing
temperature.

\begin{figure*}[htb]
\centering
\includegraphics{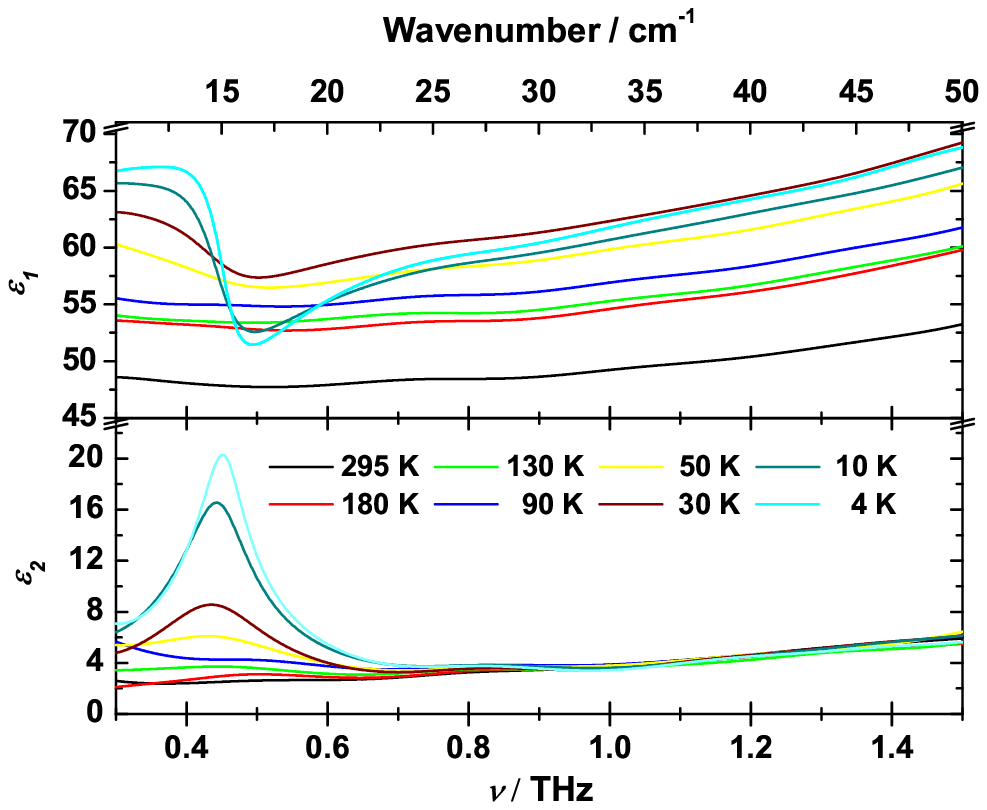}
\caption{\label{THz}(Color online) Real (upper panel) and imaginary
(lower panel) parts of the complex dielectric constant of \tb\ in
the THz regime. The color legend holds for both panels.}
\end{figure*}

Using $E_{g}$ and $E_{g}$, $A_{2g}$ and $A_{1g}$ irreducible
representations\cite{gin00,malkin04} for the ground state and
excited CF levels, respectively, it can be confirmed that indeed all
these ground state to excited state transitions are expected to be
Raman active. Through analogous considerations, the lowest
excitation ($E_{g}$ $\rightarrow$ $E_{g}$, 13 cm$^{-1}$) is found to
be symmetry forbidden in a direct dipole transition. In this
respect, terahertz time domain spectroscopy (THz TDS) in the range
of 10 to 50 cm$^{-1}$ (0.3 to 1.5 THz) was employed to probe the
absorption of \tb\ at various temperatures. Using this technique it
is possible to extract complex optical quantities in the THz range
through the direct measurement of the time trace of the transmitted
THz pulse. The obtained curves for the real ($\epsilon_{1}(\nu)$)
and imaginary ($\epsilon_{2}(\nu)$) parts of the dielectric constant
at various temperatures are plotted in fig. \ref{THz}. The plots
clearly show an absorption around 0.45 THz (15 cm$^{-1}$) at low
temperatures, corresponding to the 13 cm$^{-1}$ CF level also
observed in the Raman spectra of \tb. This is corroborated by the
identical temperature dependence of the mode in both methods: its
spectral signature decreases with increasing temperature, vanishing
around 90 K. The fact that the 13 cm$^{-1}$ CF level is observed in
a direct dipole transition indicates that this level is not a true
$E_{g}$ $\rightarrow$ $E_{g}$ transition, since such a dipole
transition would be symmetry forbidden.

Overall, the existence of the additional CF levels in \tb\,
unaccounted for by CF calculations, suggests the presence of a
second Tb$^{3+}$ site in the structure, with an energy level scheme
different from those reported previously. Three additional CF levels
are observed experimentally, while four would naively be expected
for a slightly differing Tb site in this low wavenumber region. A
fourth new CF level might be unresolved due to the strong CF level
at 83 cm$^{-1}$ or might simply be symmetry forbidden in a Raman
transition. Moreover, the fact that the $\sim$ 13 cm$^{-1}$ CF level
is simultaneously Raman and dipole active indicates the breaking of
inversion symmetry in the system, which questions the validity of
its current symmetry interpretation. Recently, the exact symmetry of
the \tb\ lattice has been extensively studied. Han \emph{et
al.}\cite{han04} performed neutron powder diffraction and x-ray
absorption fine-structure experiments down to 4.5 K, revealing a
perfect pyrochlore lattice, within experimental error. Ofer \emph{et
al.}\cite{ofer07} find no static lattice distortions on the
timescale of 0.1 $\mu$s, down to 70 mK. Most recently, however, Ruff
\emph{et al.}\cite{ruff07} found finite structural correlations at
temperatures below 20 K, indicative of fluctuations above a very low
temperature structural transition. The present experimental results
suggest these fluctuations may even induce a minute static disorder,
resulting in the observed CF level diagram.

In \dy\ the Dy$^{3+}$ ions (4$f^{9}$) have a $^{6}H_{15/2}$ ground
state. Crystal field calculations have been performed by Jana
\textit{et al.}\cite{jana02}, who deduced an energy level scheme
consisting of eight Kramers' doublets, with a separation of $\simeq$
100 cm$^{-1}$ of the first excited state. Malkin \textit{et
al.}\cite{malkin04} and Rosenkranz \textit{et
al.}\cite{rosenkranz00} estimated the first excited state gap to be
$>$ 200 cm$^{-1}$ and $\simeq$ 266 cm$^{-1}$, respectively. The
Raman spectrum of \dy\ shows only one extra excitation compared to
the 'lattice template' spectrum of \gd, at an energy of $\sim 287$
cm$^{-1}$. This excitation is tentatively ascribed to the first
excited CF level, comparing most favorably to the estimation of
Rosenkranz \textit{et al.}.

    For \ho, ground state $^{5}I_{8}$ (Ho$^{3+}$,4$f^{10}$), several
CF energy level schemes have been
calculated\cite{rosenkranz00,malkin04,siddharthan99,jana00}, all
with first excited state separations around 150 cm$^{-1}$. However,
the Raman spectrum of \ho\ does not show any clear inelastic light
scattering from CF levels to compare these calculations to. Although
there is some weak intensity around $\simeq$ 150 cm$^{-1}$, which
compares favorably with all of the calculated level schemes, the
intensity of this scattering is insufficient to definitively ascribe
it to a CF level.

\section{Conclusions}
To summarize, several members of the rare earth titanates family \R\
were studied using magnetic susceptibility measurements and
polarized inelastic light scattering experiments. Lattice
excitations were found to vary only slightly between crystals with
different rare earth ions. Temperature dependent measurements also
revealed completely analogous behavior of the phononic excitations,
except for two anomalous phonons in \tb, which seem to be coupled to
the low energy crystal field excitations in that compound. Such
crystal field excitations were observed in \tb\ and possibly also in
\dy. Only one non-phononic excitation is clearly observed in \dy,
its energy consistent with estimates of the first excited crystal
field level. For \tb, all of the the previously determined crystal
field energy levels were confirmed. Moreover, the resulting energy
level diagram was expanded by three newly observed CF levels, only
one of which has been calculated before. Also, one previously
reported level was found to be both Raman and dipole active,
contradicting its current presumed symmetry. These findings may
reflect the existence of two inequivalent Tb sites in the low
temperature structure or a static symmetry reduction of another
nature, suggesting the recently found structural correlations to
induce a minute static disorder in \tb\ at very low temperatures.
The new crystal field information may serve to help elucidate the
complex theoretical enigma of \tb.

\begin{acknowledgments}
The authors would like to thank F. van der Horst for his help using
the PW 1710 diffractometer. We also acknowledge fruitful discussions
with M. Mostovoi, D. Khomskii and S. Singh. This work is part of the
research programme of the 'Stichting voor Fundamenteel Onderzoek der
Materie (FOM)', which is financially supported by the 'Nederlandse
Organisatie voor Wetenschappelijk Onderzoek (NWO)'
\end{acknowledgments}


\begin{thebibliography}{77}
\bibitem[{ice()}]{ice}
\bibinfo{note}{The ground state configuration of the spin ice model is formed
  by satisfying conditions analogous to Pauling's 'ice rules': every
  tetrahedron has both two spins pointing into and two spins pointing out of
  the tetrahedron. This (under)constraint on the system again results in a
  macroscopically large number of degenerate ground states.}
\end{thebibliography}
\end{document}